%% file: menu.tex
\documentclass[twocolumn]{aastex63}
\usepackage{bm}
\usepackage{soul}
\usepackage{xfrac}
\usepackage{xcolor}
\usepackage{amsmath}
\usepackage{CJKutf8}
\usepackage{hyperref}
\usepackage{textcomp}
\usepackage{multirow}
\usepackage{graphicx}
\usepackage{booktabs}
\usepackage{tabularx}
\usepackage{rotating}
\usepackage{makecell}
\usepackage{mathrsfs}
\usepackage{ragged2e}
\usepackage{nicefrac}
\usepackage{mathtools}
\usepackage[T1]{fontenc}
\usepackage[utf8]{inputenc}
\usepackage[normalem]{ulem}
\usepackage[makeroom]{cancel}

\newcommand{\V}{\textit{\textbf{V}}}
\newcommand{\W}{\textit{\textbf{W}}}

\newcommand{\tworows}[1]{\multirow{2}{*}{#1}}

\newcommand{\simpchinese}[1]{\begin{CJK}{UTF8}{gbsn}#1\end{CJK}}
\newcommand{\tradchinese}[1]{\begin{CJK}{UTF8}{bsmi}#1\end{CJK}}

\defcitealias{bi_puffed-up_2021}{BLD21}

\begin{document}

\title{Gap-opening Planets Make Dust Rings Wider}

\author[0000-0002-0605-4961]{Jiaqing Bi \simpchinese{(毕嘉擎)}}
\affil{Department of Physics \& Astronomy, University of Victoria, 3800 Finnerty Road, Victoria, BC V8P 5C2, Canada}
\affil{Academia Sinica Institute of Astronomy \& Astrophysics, No. 1, Sec. 4, Roosevelt Rd, Taipei 10617, Taiwan}
\email{bijiaqing@uvic.ca}

\author[0000-0002-8597-4386]{Min-Kai Lin \tradchinese{(林明楷)}}
\affil{Academia Sinica Institute of Astronomy \& Astrophysics, No. 1, Sec. 4, Roosevelt Rd, Taipei 10617, Taiwan}
\affil{Physics Division, National Center for Theoretical Sciences, Taipei 10617, Taiwan}
\email{mklin@asiaa.sinica.edu.tw}

\author[0000-0001-9290-7846]{Ruobing Dong \simpchinese{(董若冰)}}
\affil{Department of Physics \& Astronomy, University of Victoria, 3800 Finnerty Road, Victoria, BC V8P 5C2, Canada}
\email{rbdong@uvic.ca}


\begin{abstract}
    
    As one of the most commonly observed disk substructures, dust rings from high-resolution disk surveys appear to have different radial widths. Recent observations on PDS 70 and AB Aur reveal not only planets in the disk, but also the accompanying wide dust rings. We use three-dimensional dust-and-gas disk simulations to study whether gap-opening planets are responsible for the large ring width in disk observations. We find that gap-opening planets can widen rings of dust trapped at the pressure bump via planetary perturbations, even with the mid-plane dust-to-gas ratio approaching order unity and with the dust back-reaction accounted for. We show that the planet-related widening effect of dust rings can be quantified using diffusion-advection theory, and provide a generalized criterion for an equilibrated dust ring width in three-dimensional disk models. We also suggest that the ring width can be estimated using the gas turbulent viscosity $\alpha_{\rm turb}$, but with cautions about the Schmidt number greater than order unity.

\end{abstract}

\keywords{Protoplanetary disks (1300); Planet formation (1241); Circumstellar dust (236); Astrophysical dust processes (99); Astronomical simulations (1857)}


\input{Section_1}
\input{Section_2}
\input{Section_3}
\input{Section_4}
\input{Section_5}

\clearpage
\bibliography{jiaqing}


\end{document}

%% file: Section_1.tex
\section{Introduction} \label{sec:intro}

Facilitating radio interferometers like the Atacama Large Millimeter/submillimeter Array (ALMA), many dust rings and gaps (e.g., HL Tau, \citealt{alma_partnership_2014_2015}; TW Hya, \citealt{huang_co_2018}), as well as dust spirals \citep[e.g., MWC 758,][]{dong_eccentric_2018} and lopsided dust clumps \citep[e.g., IRS 48,][]{van_der_marel_major_2013}, have been revealed in protoplanetary disks (PPDs). Understanding the morphology of those detailed disk structures is critical to the topic of planet formation, as they could be the indicator of on-going planet-disk interactions \citep{zhang_disk_2018, hammer_which_2021}.

\begin{figure}[tp]
\includegraphics[width = 0.47\textwidth]{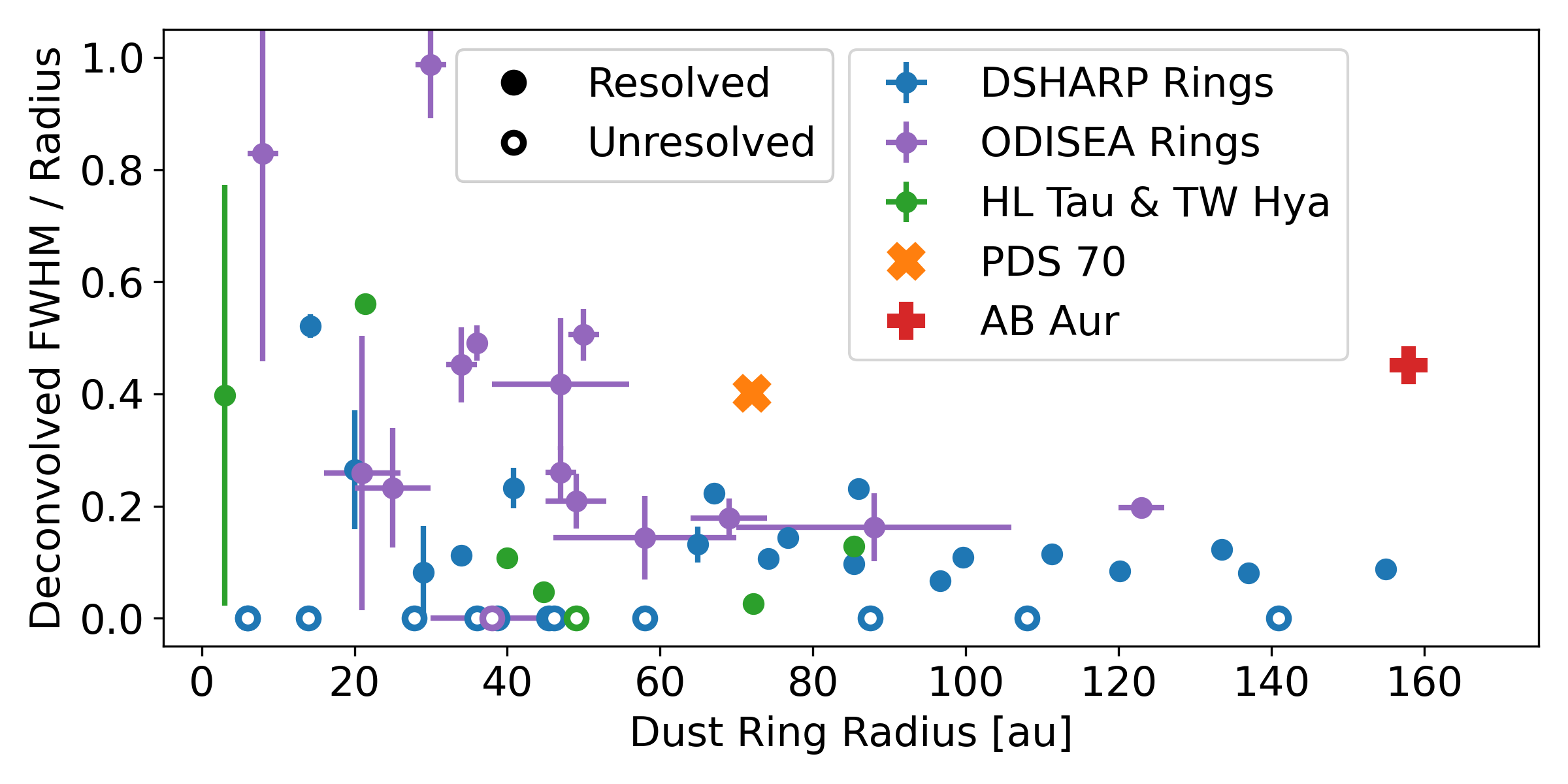}
\figcaption{
The ratio between the deconvolved full width at half maximum (FWHM) and the radius of observed dust rings. The deconvolution method can be found in \cite{dullemond_disk_2018}, with unresolved widths set to zero (open circles in the plot). Samples are selected from the DSHARP Program \citep{andrews_disk_2018,huang_dsharp2_2018}, the ODISEA Program \citep{cieza_ophiuchus_2021}, HL Tau \citep{alma_partnership_2014_2015}, TW Hya \citep{huang_co_2018}, PDS 70 \citep{benisty_circumplanetary_2021,portilla-revelo_self-consistent_2022}, and AB Aur \citep{tang_planet_2017}. Rings without errorbars in the DSHARP Program are not sampled.
\label{fig:wrratio}}
\end{figure}

\begin{figure*}[tp]
\includegraphics[width = \textwidth]{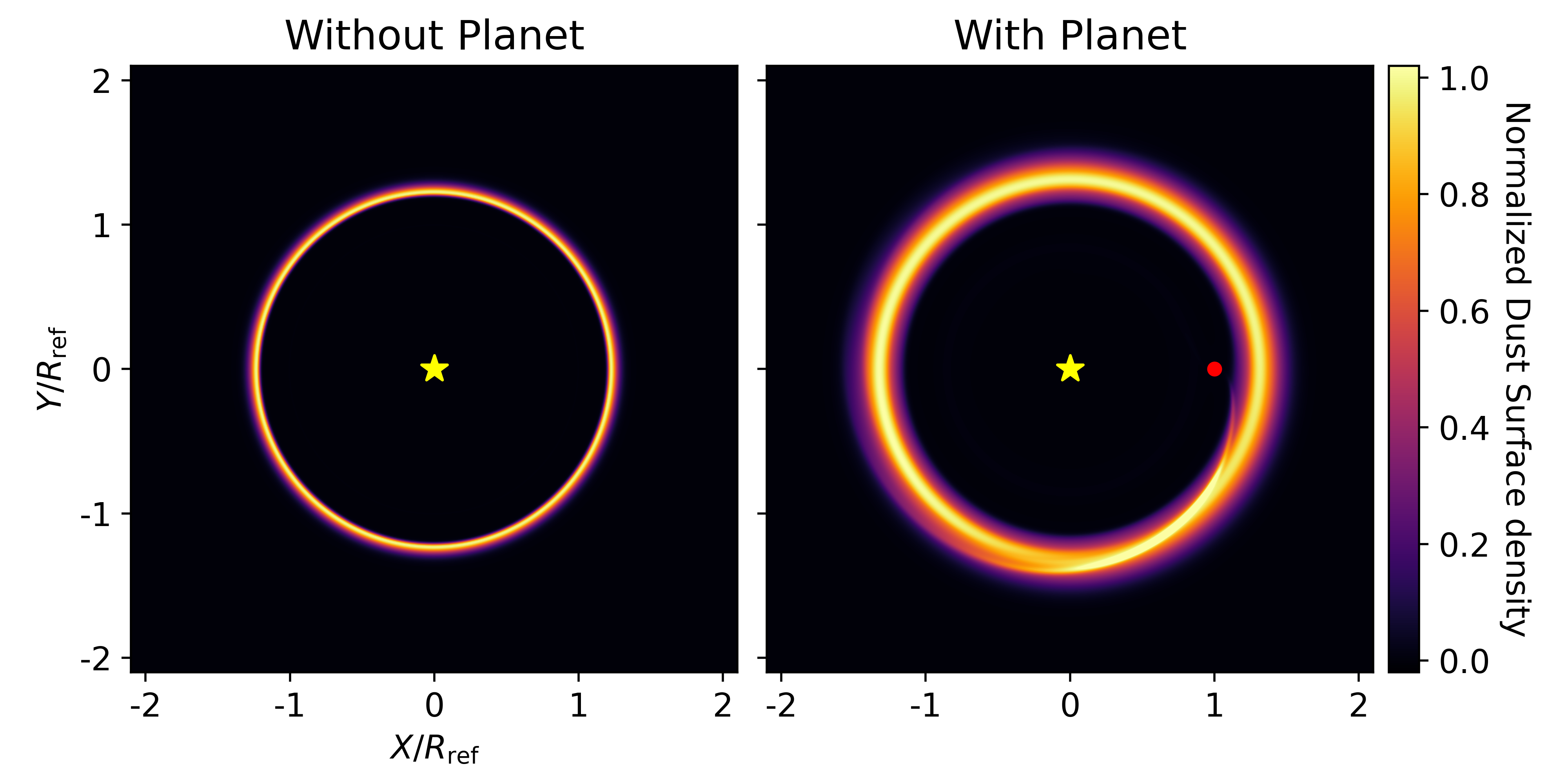}
\figcaption{The face-on view of dust rings trapped at the pressure bump. The left panel shows a dust ring in a disk model with no planet (Model C1), and the right panel shows a dust ring in the disk with a Saturn-mass gap-opening planet (Model C5, see Section \ref{sec:model_setup} and \ref{sec:resultc} for descriptions and applications). Both snapshots are taken at $t = 1000T_{\rm ref}$ in each model, and are normalized individually by the radial peak in the azimuthally averaged profile. The radial profiles of dust surface density are shown in Figure \ref{fig:modelc}. The yellow star indicates the location of the central star and the red dot indicates the location of the planet. 
\label{fig:ring_compare}}
\end{figure*}

As one of the most commonly observed substructures \citep{andrews_disk_2018, long_gaps_2018, van_der_marel_protoplanetary_2019}, dust rings have different radial widths, 
with the ratio between the deconvolved full width at half maximum (FWHM) and the radius ranging from $\lesssim 10\%$ \citep[e.g., AS 209,][]{huang_dsharp2_2018} to $\gtrsim 90\%$ \citep[e.g., EM* SR 24S,][]{cieza_ophiuchus_2021} (see Figure \ref{fig:wrratio}). This wide range of widths indicates different relative strengths of dust trapping (e.g., gas drag in the pressure bump, \citealt{takeuchi_radial_2002, paardekooper_planets_2004}; radial traffic jam due to aggregate sintering, \citealt{okuzumi_sintering-induced_2016}) and dust diffusion (e.g., turbulence-induced diffusion, \citealt{youdin_particle_2007}).

Gap-opening planets have now become a popular explanation to the formation of dust rings, since pressure bumps naturally arise from the gap-opening process \citep{lin_nonaxisymmetric_1993}. A great example of this scenario could be PDS 70, where two planets have been detected in the gap next to a dust ring \citep{keppler_discovery_2018, haffert_two_2019, benisty_circumplanetary_2021}. In the meantime, this dust ring is visibly wider compared with many other well-resolved rings from disk surveys (see Figure \ref{fig:wrratio}). Similar things happen to AB Aur, where a planet is also detected next to a wide dust ring \citep{tang_planet_2017, currie_images_2022}. We therefore question whether this large ring width is associated with planets, and how much, if associated, the planet can widen the dust ring.

To answer those, we perform three-dimensional grid-based hydrodynamic simulations of dusty PPDs with embedded gap-opening planets. We find that the planet can widen dust rings trapped at the pressure bump (see Figure \ref{fig:ring_compare}). We also provide an analytical approach to estimate the dust ring width in such a scenario. Our results suggest that gap-opening planets could be a potential explanation to wide dust rings in PPDs.

This paper is organized as follows. We describe the disk--planet system of interest and different numerical models in Section \ref{sec:model}. We present our results in Section \ref{sec:result}, which includes three parts: The first part shows the widening effect of dust rings due to planet-related effects. The second part shows the combined effect of the planet and different levels of dust back-reaction on dust rings. The third part provides an analytical approach to estimate the dust ring width. In Section \ref{sec:discussion} we discuss the connection between dust diffusivity and gas turbulent viscosity under planet-related effects. Finally, we conclude in Section \ref{sec:conclusion}.

%% file: Section_2.tex
\section{Planet-Disk Model} \label{sec:model}

We consider a 3D protoplanetary disk composed of gas and dust with an embedded planet of mass $M_{\rm p}$ around a central star of mass $M_\star$. We neglect disk self-gravity, magnetic fields, planet orbital migration, and planet accretion. We use $\{r, \phi, \theta\}$ to denote spherical radius, azimuth, and polar angle, and $\{R, \phi, Z\}$ to denote cylindrical radius, azimuth, and height. Both coordinates are centered on the star. 

In the following sections, the subscript ``ref'' denotes azimuthally averaged values at $R = R_{\rm ref}$ on the disk midplane, where $R_{\rm ref}$ is a reference radius. The subscript ``0'' is only used for time-varying quantities, and it denotes their initial values. And the symbol $\langle\rangle$ denotes azimuthally averaged values.

\subsection{Basic Equations and Numerical Setups}

Models in this paper use the same basic equations and numerical setups as those in \cite{bi_puffed-up_2021} (hereafter \citetalias{bi_puffed-up_2021}). Therefore, we only briefly list them here and refer readers to \citetalias{bi_puffed-up_2021} for detailed descriptions and justifications.

The volumetric density, pressure, and velocity of gas are denoted by $\{\rho_{\rm g}, P, \V\}$. The time-independent, vertically isothermal, axisymmetric gas temperature and sound speed are given by $T(R) = T_{\rm ref}(R/R_{\rm ref})^{-q}$ and $c_{\rm s}(R) = c_{\rm s, ref}(R/R_{\rm ref})^{-q/2}$. The isothermal equation of state and the pressure scale-height are given by $P = \rho_{\rm g}c_{\rm s}^2$ and $H_{\rm g} = c_{\rm s}/\Omega_{\rm K}$, where $\Omega_{\rm K}(R) = \sqrt{GM_\star/R^3}$ is the Keplerian angular velocity and $G$ is the gravitational constant. The disk is assumed non-flared with a constant aspect ratio $h = H_{\rm g}/R = 0.05$, corresponding to $q = 1$.

We consider a single species of dust modeled as a pressureless fluid with volumetric density and velocity $\{\rho_{\rm d}, \W\}$. The Epstein gas drag on the dust is parametrized by the Stokes number ${\rm St} = \tau_{\rm s}\Omega_{\rm K}$, where 
\begin{equation} \label{eq:taus}
    \tau_{\rm s} = 
    \frac{\rho_{\rm g,norm}}{\rho_{\rm g}}\frac{c_{\rm s,ref}}{c_{\rm s}}\frac{{\rm St}_{\rm 0,ref}}{\Omega_{\rm K,ref}}
\end{equation}
is the particle stopping time. $\rho_{\rm g,norm}$ is a normalization that equals to $\rho_{\rm g0,ref}$ in type A models (see Section \ref{sec:model_setup}).

The hydrodynamic equations for gas and dust are given by 
\begin{align}
     \label{eq:gascont}
    &\frac{\partial \rho_{\rm g}}{\partial t} + \nabla \cdot (\rho_{\rm g} \V) = 0, \\
    &\begin{aligned}
        \frac{\partial \V}{\partial t} + \V \cdot \nabla \V = &- \frac{1}{\rho_{\rm g}} \nabla P - \nabla \Phi \\
        &+ \frac{\epsilon_\rho}{\tau_{\rm s}}(\W - \V) + \frac{1}{\rho_{\rm g}} \nabla \cdot \mathcal{T},
     \end{aligned} \\
     \label{eq:dustcont}
    &\frac{\partial \rho_{\rm d}}{\partial t} + \nabla \cdot (\rho_{\rm d} \W) = 0, \\ 
    &\frac{\partial \W}{\partial t} + \W \cdot \nabla \W =  - \nabla \Phi - \frac{1}{\tau_{\rm s}}(\W - \V).
\end{align}
Here $\Phi = -GM_\star/r + \Phi_{\rm p} + \Phi_{\rm ind}$ is the net gravitational potential composed of terms from the star, the planet, and the indirect planet–star gravitational interactions, respectively. The disk-related potential terms are neglected for the non-self-gravitating disk. $\epsilon_\rho = \rho_{\rm d}/\rho_{\rm g}$ is the \textit{local} dust-to-gas ratio, which should be distinguished from the vertically integrated \textit{global} dust-to-gas ratio $\epsilon_\Sigma = \Sigma_{\rm d}/\Sigma_{\rm g}$. $\mathcal{T}$ is the viscous stress tensor (see Equation 11 in \citetalias{bi_puffed-up_2021}) which involves a gas kinematic viscosity $\nu$. We implement $\nu = 10^{-5} R_{\rm ref}^2 \Omega_{\rm K,ref}$ to suppress interference such as vertical shear instability \citep{nelson_linear_2013} and vortex formation \citep{koller_vortices_2003, li_potential_2005, li_type_2009, lin_type_2010} at gap edges. To isolate planet-related effects on the dust ring, turbulent-induced dust diffusion \citep{weber_predicting_2019} is neglected. Unless otherwise specified, dust back-reaction on the gas is included. 

We consider a planet on a fixed, circular orbit at $R = R_{\rm ref}$ on the disk midplane. The planet-related potential terms are 
\begin{equation}
    \Phi_{\rm p} + \Phi_{\rm ind} = - \frac{Gm_{\rm p}(t)}{\sqrt{r^{\prime2} + r_{\rm s}^2}} + \frac{Gm_{\rm p}(t)}{R_{\rm ref}^2}R\cos{(\phi - \phi_{\rm p})},
\end{equation}
where $\phi_{\rm p}$ is the azimuth of the planet, $r_{\rm s} = 0.1H_{\rm g}$ is a smoothing length, and $r^\prime$ is the distance to the planet. Here we define a time-dependent planet mass $m_{\rm p}(t)$, with its value increasing gradually from zero to $M_{\rm p}$ (see Equation 14 in \citetalias{bi_puffed-up_2021}) at the start of simulations to avoid transient impacts of adding a full-mass planet to the disk. The planet's potential is turned on over a timescale of $t_{\rm g} = 500T_{\rm ref}$, where $T_{\rm ref} = 2\pi\Omega_{\rm K,ref}^{-1}$ is the reference orbital period.

Our models are evolved by the FARGO3D code \citep{benitez-llambay_fargo3d_2016, benitez-llambay_asymptotically_2019}. We adopt a spherical domain centered on the star with $r \in [0.2, 4.0]R_{\rm ref}$, $\phi \in [0, 2\pi]$, and polar angle such that $\tan(\pi/2 - \theta) \in [-3h, 3h]$. The resolutions we choose are $N_{r} \times N_\theta \times N_\phi = 360 \times 90 \times 720$, with logarithmic spacing in $r$ and uniform spacing in $\theta$ and $\phi$. 

The gas density is damped to its initial value at radial boundaries, and is assumed to be in vertical hydrostatic equilibrium at vertical boundaries. The dust density is symmetric at both radial and vertical boundaries. The meridional velocities of gas and dust are set to zero at radial and vertical boundaries, except that the inner radial boundary is open for mass loss of dust. The azimuthal velocities at those boundaries are assigned at the Keplerian speed with a pressure offset for gas. Periodic boundaries are imposed in the $\phi$ direction.

\subsection{Models} \label{sec:model_setup}

The models in our paper are categorized into three types, namely type A, B, and C. The three types differ in the prescription of gas evolution and initial conditions. In type A models, the initial radial profile of dust surface density $\Sigma_{\rm d0}$ is a power-law function, whereas in type B and C models $\Sigma_{\rm d0}$ is a Gaussian radial bump. Different from that in type B models, the evolution of $\{\rho_{\rm g}, \V\}$ is artificially stalled in C, which means they are time-invariant.

\subsubsection{Type A Models: How will $M_{\rm p}$ and St affect the widening effect?} \label{sec:typea}

Type A models are used to \textit{quantitatively} study how planet-related effects would change the radial width of dust rings at the outer gap edge. The corresponding results are shown in Section \ref{sec:resulta} and \ref{sec:discussion}. There are twelve type A models with different planet masses $M_{\rm p}$ ranging from $2\times10^{-4}M_\star$ to $7\times10^{-4}M_\star$, and two initial reference Stokes numbers ${\rm St}_{\rm 0,ref}=10^{-3}$ and $10^{-2}$. Here we define the model with $\{M_{\rm p}, {\rm St}_{\rm 0,ref}\} = \{3\times10^{-4}M_\star, 10^{-3}\}$ as the fiducial model, representing a Saturn-mass planet around a solar-mass star, with 0.1-millimeter-sized grains at $\sim 45$ au in a young PPD such as HL Tau\footnote{Here we assume that the grain internal density is 1.5 g/cm$^3$, the total disk mass is 0.2 $M_\odot$ \citep{booth_13c17o_2020}, the outer disk radius is 150 au, and the surface density power-law index is -1.5.}.

The initialization of gas and dust in type A models are the same as those in \citetalias{bi_puffed-up_2021}. The axisymmetric gas density profile is initialized to 
\begin{equation} \label{eq:rhog}
    \rho_{\rm g0} = \rho_{\rm g0,ref}\left(\frac{R}{R_{\rm ref}}\right)^{-p} 
    \times \exp\left[\frac{GM_\star}{c_{\rm s}^2}\left(\frac{1}{r} - \frac{1}{R}\right)\right],
\end{equation}
with $p = 1.5$ and $\rho_{\rm g0,ref}$ being arbitrary for a non-self-gravitating disk. The dust density is initialized to 
\begin{equation}
    \rho_{\rm d0} = \bigg(\epsilon_{\rho{\rm 0}}\rho_{\rm g0}\bigg)\bigg\vert_{Z=0} \times \exp{\left(-\frac{Z^2}{2H_{\epsilon}^2}\right)},
\end{equation}
where $\epsilon_{\rho0}\vert_{Z=0} = 0.1$ and $H_{\epsilon} = H_{\rm g}H_{\rm d}(H_{\rm g}^2 - H_{\rm d}^2)^{-1/2}$ is valued such that the dust scale-height $H_{\rm d0} = 0.1H_{\rm g}$ and $\epsilon_{\Sigma0} = 0.01$. The azimuthal velocities are initialized to 
\begin{align}
    V_{\phi 0} = & \; R\Omega_{\rm K}\left(\sqrt{1-2\eta} + 
        \frac{\epsilon_{\rho0}\eta}{\epsilon_{\rho0}+1}\frac{1}{{\rm St}^{\prime2}+1}\right) \\
    W_{\phi 0} = & \; \sqrt{\frac{GM_\star}{r}} - 
        R\Omega_{\rm K}\left(\frac{\eta}{\epsilon_{\rho0}+1}\frac{1}{{\rm St}^{\prime2}+1}\right),
\end{align}
where ${\rm St}^{\prime} = {\rm St}/(1+\epsilon_\rho)$, and $2\eta = (p+q)h^2 + q (1 - R/r)$ is a dimensionless measurement of the radial pressure gradient. The radial velocities are initialized to
\begin{alignat}{2}
    V_{R0} = & &&\frac{2\epsilon_{\rho0}\eta}{\epsilon_{\rho0}+1}
        \frac{{\rm St}^{\prime}}{{\rm St}^{\prime2}+1}R\Omega_{\rm K} \\
    W_{R0} = & - &&\frac{2\eta}{\epsilon_{\rho0}+1}
        \frac{{\rm St}^{\prime}}{{\rm St}^{\prime2}+1}R\Omega_{\rm K}.
\end{alignat}
And the initial vertical velocities are $V_{Z0} = W_{Z0} = 0$. 

\subsubsection{Type B Models: How will dust back-reaction affect the widening effect?} \label{sec:typeb}

Type B models are used to \textit{qualitatively} study the effect of different levels of dust back-reaction on the dust ring width. The corresponding results are shown in Section \ref{sec:resultb} and Figure \ref{fig:modelb}. The models include a gas gap already opened by a planet, and a Gaussian dust ring at the outer gap edge. There are five type B models, namely B1 to B5, with different levels of dust load in the dust ring.

To have a gas gap opened by the planet, the gas initialization $\{\rho_{\rm g0}, \V_0\}$ in all type B models are taken from the snapshot at $t = 3000 T_{\rm ref}$ in the fiducial type A model (i.e., $\{M_{\rm p}, {\rm St}_{\rm 0,ref}\} = \{3\times10^{-4}M_\star, 10^{-3}\}$), in which the gap profile has almost reached an equilibrium state. Here we set $t_{\rm g} = 0$ to maintain the initial gap profile, and consequently, the planet mass is fixed at $m_{\rm p} = 3\times10^{-4}M_\star$. 

Dust grains in type B models have ${\rm St}_{\rm 0,ref} = 10^{-3}$. The dust density is initialized to an axisymmetric Gaussian ring with
\begin{equation}
    \rho_{\rm d0} = \frac{\Sigma_{\rm d0}}{\sqrt{2\pi}H_{\rm d0}} 
    \times \exp{ \left[ -\frac{(R-R_{\rm pmax})^2}{2w_{\rm d0}^2} - \frac{Z^2}{2H_{\epsilon}^2} \right]},
\end{equation}
where $H_{\rm d0} = 0.1H_{\rm g}$, $R_{\rm pmax} = 1.243R_{\rm ref}$ is the radius of the midplane pressure maximum, and $w_{\rm d0} = H_{\rm g,ref}$ is the initial radial width of the ring. $\Sigma_{\rm d0}$ is valued such that $\epsilon_{\Sigma0} = \{5\times10^{-4}, 5\times10^{-3}, 5\times10^{-2}, 5\times10^{-1}, 5\}$ at $R = R_{\rm pmax}$ in Model B1, B2, B3, B4, B5, respectively (see Table \ref{tab:modelb}). We note that $\epsilon_{\Sigma0}$ for Model B3 best reproduces the dust load level of the ring at $t = 3000 T_{\rm ref}$ in the fiducial type A model. The dust velocities in all type B models are initialized to $W_{\phi0} = \sqrt{GM_\star/r}$ and $W_{R0} = W_{Z0} = 0$.

\input{table_modelb}

\subsubsection{Type C Models: Isolating the multiple effects caused by the planet} \label{sec:typec}

Type C models are used to \textit{qualitatively} study how various planet-related effects would change the dust ring width. The corresponding results are shown in Section \ref{sec:resultc} and Figure \ref{fig:modelc}. Like type B models, these models also include a Gaussian dust ring and a gas gap, but the further evolution of gas density and velocity is stalled to control variables. 

There are five type C models to isolate planet-related effects, which are differentiated by the initialization of gas and whether planetary potential terms are included (see Table \ref{tab:modelc}). The face-on view of gas initialization in Model C1 and C5 are provided in Appendix \ref{app:eclip} to visualize the differences. The initialization of dust in all type C models are identical to that in Model B3 (i.e., an axisymmetric Gaussian ring with $\epsilon_{\Sigma0} = 0.05$ at $R = R_{\rm pmax}$ and ${\rm St}_{\rm 0,ref} = 10^{-3}$). In Model C1, the prescription of the planet is identical to the one in type B models. To avoid errors while disabling gas evolution, the gas damping effect at boundaries and dust back-reaction on the gas are disabled as well.

\input{table_modelc}

%% file: table_modelb.tex
\begin{table}[t]

\centering
\caption{Dust Load Initialization of Type B Models} \label{tab:modelb}

\medskip
\begin{tabular*}{0.47\textwidth}{cc}
\toprule \\[-1.2em]

\hspace*{0.08\textwidth} Model                          \hspace*{0.08\textwidth} & 
\hspace*{0.08\textwidth} $\epsilon_{\Sigma0}^{\rm max}$ \hspace*{0.08\textwidth} \\ \\[-1em]

\midrule \\[-1.2em]

B1 & 0.0005 \\ \\[-1.2em]

B2 & 0.005 \\ \\[-1.2em]

B3 & 0.05 \\ \\[-1.2em]

B4 & 0.5 \\ \\[-1.2em]

B5 & 5 \\ \\[-1.2em]

\bottomrule
\end{tabular*}

\justify
\tablecomments{$\epsilon_{\Sigma0}^{\rm max}$ is the initial dust-to-gas surface density ratio at the radial peak of the dust ring. Except for the dust load, all the other dust and gas initial conditions are identical for all type B models.}

\end{table}

%% file: table_modelc.tex
\begin{table*}[t]

\centering
\caption{Gas Initialization of Type C Models} \label{tab:modelc}

\medskip
\begin{tabular*}{\textwidth}{@{\extracolsep{\fill}}ccccccc}
\toprule \\[-1.2em]

Model 
    & $\rho_{\rm g}$ 
    & $V_{\rm R}$ 
    & $V_{\phi}$ 
    & $V_{\rm Z}$ 
    & $\Phi_{\rm p} + \Phi_{\rm ind}$ 
    & Axisymmetry \\ \\[-1em]

\midrule \\[-1.2em]

\tworows{C1} 
    & \tworows{$\rho_{\rm g}^\prime$} 
    & \tworows{$V_{\rm R}^\prime$} 
    & \tworows{$V_{\phi}^\prime$} 
    & \tworows{$V_{\rm Z}^\prime$} 
    & \tworows{Included} 
    & with density spirals\\ \\[-1.5em]
    &&&&&& with velocity spirals  \\ \\[-0.75em]

\tworows{C2} 
    & \tworows{$\rho_{\rm g}^\prime$}
    & \tworows{$V_{\rm R}^\prime$}
    & \tworows{$V_{\phi}^\prime$}
    & \tworows{$V_{\rm Z}^\prime$}
    & \tworows{Neglected}
    & with density spirals\\ \\[-1.5em]
    &&&&&& with velocity spirals  \\ \\[-0.75em]

\tworows{C3} 
    & \tworows{$\langle\rho_{\rm g}^\prime\rangle$}
    & \tworows{$V_{\rm R}^\prime$}
    & \tworows{$V_{\phi}^\prime$}
    & \tworows{$V_{\rm Z}^\prime$}
    & \tworows{Neglected}
    & without density spirals\\ \\[-1.5em]
    &&&&&& with velocity spirals  \\ \\[-0.75em]

\tworows{C4} 
    & \tworows{$\rho_{\rm g}^\prime$}
    & \tworows{0} 
    & \tworows{$\sqrt{R^2\Omega_{\rm K}^2\left(1-\dfrac{3}{2}\dfrac{Z^2}{R^2}\right) + \dfrac{R}{\rho_{\rm g}^\prime}\dfrac{\partial P^\prime}{\partial R}}$}
    & \tworows{0}
    & \tworows{Neglected}
    & with density spirals\\ \\[-1.5em]
    &&&&&& without velocity spirals  \\ \\[-0.75em]

\tworows{C5}
    & \tworows{$\langle\rho_{\rm g}^\prime\rangle$}
    & \tworows{0}
    & \tworows{$\sqrt{R^2\Omega_{\rm K}^2\left(1-\dfrac{3}{2}\dfrac{Z^2}{R^2}\right) + \dfrac{R}{\langle\rho_{\rm g}^\prime\rangle}\dfrac{\partial \langle P^\prime\rangle}{\partial R}}$}
    & \tworows{0}
    & \tworows{Neglected}
    & without density spirals\\ \\[-1.5em]
    &&&&&& without velocity spirals  \\ \\[-1em]

\bottomrule
\end{tabular*}

\justify
\tablecomments{The superscript ``$\prime$'' in this table denotes evaluations from the snapshot at $t = 3000 T_{\rm ref}$ in the fiducial Type A model, which are asymmetric fields with planet-driven spirals and meridional flows. The symbol $\langle\rangle$ denotes azimuthally averaged values. $V_{\rm \phi}$ in model C4 and C5 assume hydrodynamic equilibrium. All values listed in this table are time-invariant in type C models. The last column explains the outcome of modifying $\{\rho_{\rm g}, \V\}$ fields. The velocity spiral refers primarily to the nonzero gas radial velocity correlated to the planet-driven spirals.}

\end{table*}

%% file: Section_3.tex
\section{Results} \label{sec:result}

\subsection{Dust Rings Widened by the Gap-opening Planet} \label{sec:resultc}

Compared with pressure bumps that are not planet-related (e.g., formed at the edge of dead zones or condensation frontiers), those at the outer edge of planet-opened gaps are additionally perturbed, as they are periodically swept by planetary wakes. In the meantime, dust grains trapped in those pressure bumps feel the wakes in two ways: perturbations in the gas density field that change the local stopping time of dust, and perturbations in the gas velocity field that change the velocity of dust. Here we show these two mechanisms, together with the insignificant non-axisymmetric planetary potential on the dust (see below), can widen the dust ring.

\begin{figure}[tp]
\includegraphics[width = 0.47\textwidth]{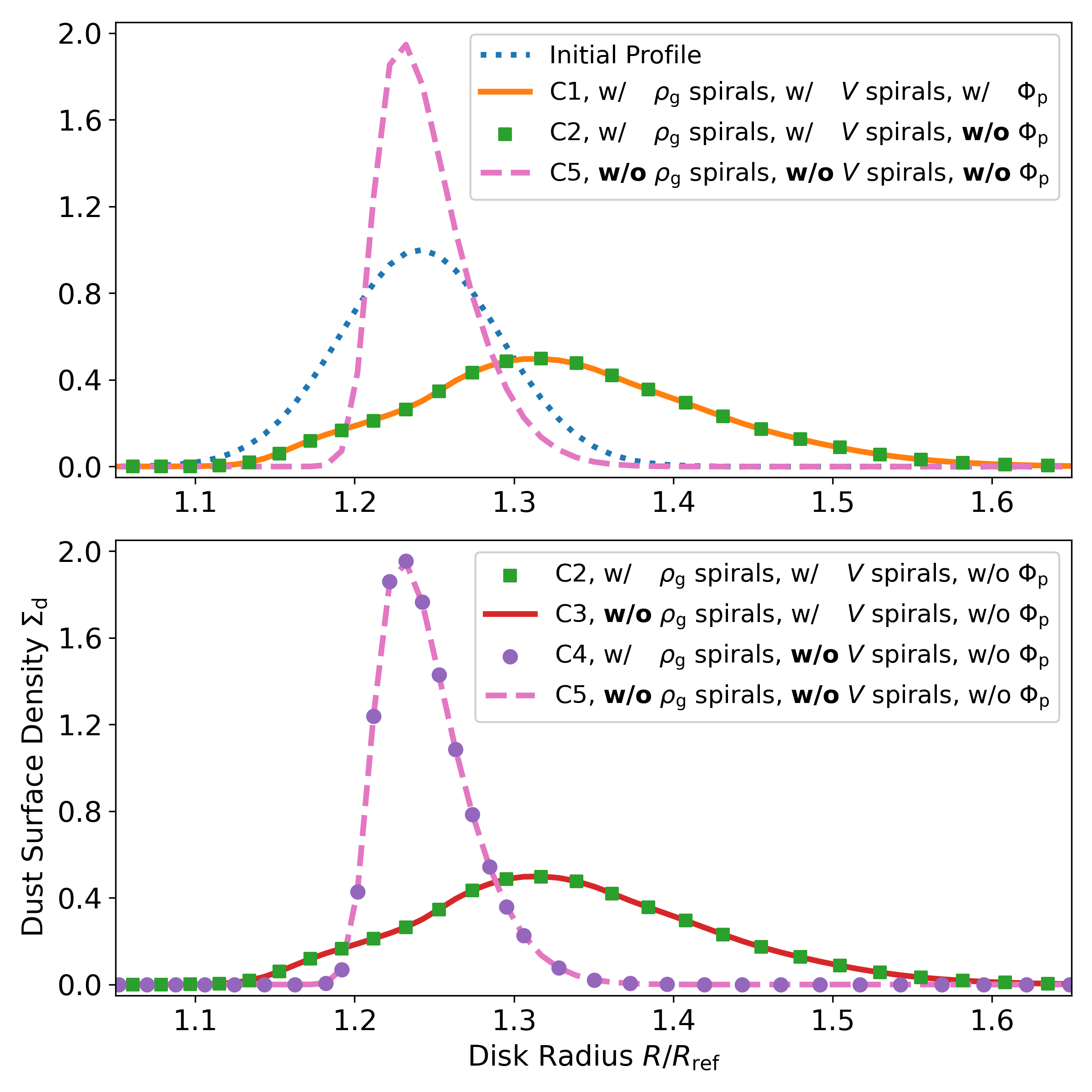}
\figcaption{
The azimuthally averaged dust surface density at $t = 1000T_{\rm ref}$ in type C models. Two panels are provided for more convenient comparisons. All profiles are normalized to the peak value in the initial profile. The initial profile peaks at the radius of the pressure maximum in the midplane. The corresponding face-on views in Model C1 and C5 are provided in Figure \ref{fig:ring_compare}. 
\label{fig:modelc}}
\end{figure}

Figure \ref{fig:modelc} shows the dust surface density profile in type C models at $t = 1000T_{\rm ref}$. We first focus on the comparison between Model C1 and C5 in the top panel, which are static-gas models including a Gaussian dust ring at the edge of a planet-opened gap, with all three planet-related effects preserved in Model C1 and eliminated in C5 (see Table \ref{tab:modelc}). Starting from the same initialization, the dust ring in Model C1 becomes much wider, whereas that in C5 continues to concentrate to the pressure maximum. This comparison shows that the net effect of the three planet-related effects widens the dust ring. 

We then question which effect of the three contributes the most to the widening effect. We isolate the three effects by turning off the planetary potential and individually modifying $\rho_{\rm g}$ and $\V$ in Model C2, C3, and C4 (see Table \ref{tab:modelc}), respectively. In the top panel of Figure \ref{fig:modelc}, the comparison between Model C1 and C2 shows that the planetary potential on the dust is insignificant. In the bottom panel, \{C2 vs C3\} and \{C4 vs C5\} show that the perturbation in the gas density field is not the dominant effect. Finally, \{C2 vs C4\} and \{C3 vs C5\} show that, it is the planetary wakes in the gas velocity field that is most responsible for the widening of the dust ring.

Meanwhile, we note that the dominance of gas velocity perturbations may only be applicable to the well-coupled dust ($\tau_{\rm s} \ll \Omega_{\rm K}^{-1}$) in our models, for which the terminal velocity approximation \citep{youdin_streaming_2005, jacquet_linear_2011, price_fast_2015, lovascio_dynamics_2019}
\begin{equation} \label{eq:approx}
\W = \V + \frac{\nabla P}{\rho_{\rm g}(1 + \epsilon_\rho)}\tau_{\rm s}
\end{equation} 
is always dominated by the gas kinematics term, even if the density perturbations launched by the planet can lead to changes of $\tau_{\rm s}$ by a few tens of percent. For larger grains in the disk, we would expect the density perturbations to play a more important role. Besides, we note that dust rings in Model C1, C2, and C3 are carried to larger radii while being widened by planet-related effects. This is the result of disabling dust back-reaction, which is required by the no-gas-evolution implementation, and will be discussed in Section \ref{sec:resultb}.   

\subsection{The Effect of Dust Back-reaction on Dust Rings} \label{sec:resultb}

For a long time, numerical disk models neglect dust back-reaction on the gas for simplicity and to reduce computational cost. This approximation worked well since conventional thoughts indicate a low dust-to-gas ratio ($\epsilon_\rho \ll 1$) in protoplanetary disks. However, \cite{kanagawa_impacts_2018} showed that at dust rings, where the dust concentrates, effective dust back-reaction is capable of flattening the radial profile of the pressure bump, leading to a broadened dust ring. Therefore, dust back-reaction is critical to the study of dust ring morphology.

\begin{figure}[tp]
\includegraphics[width = 0.47\textwidth]{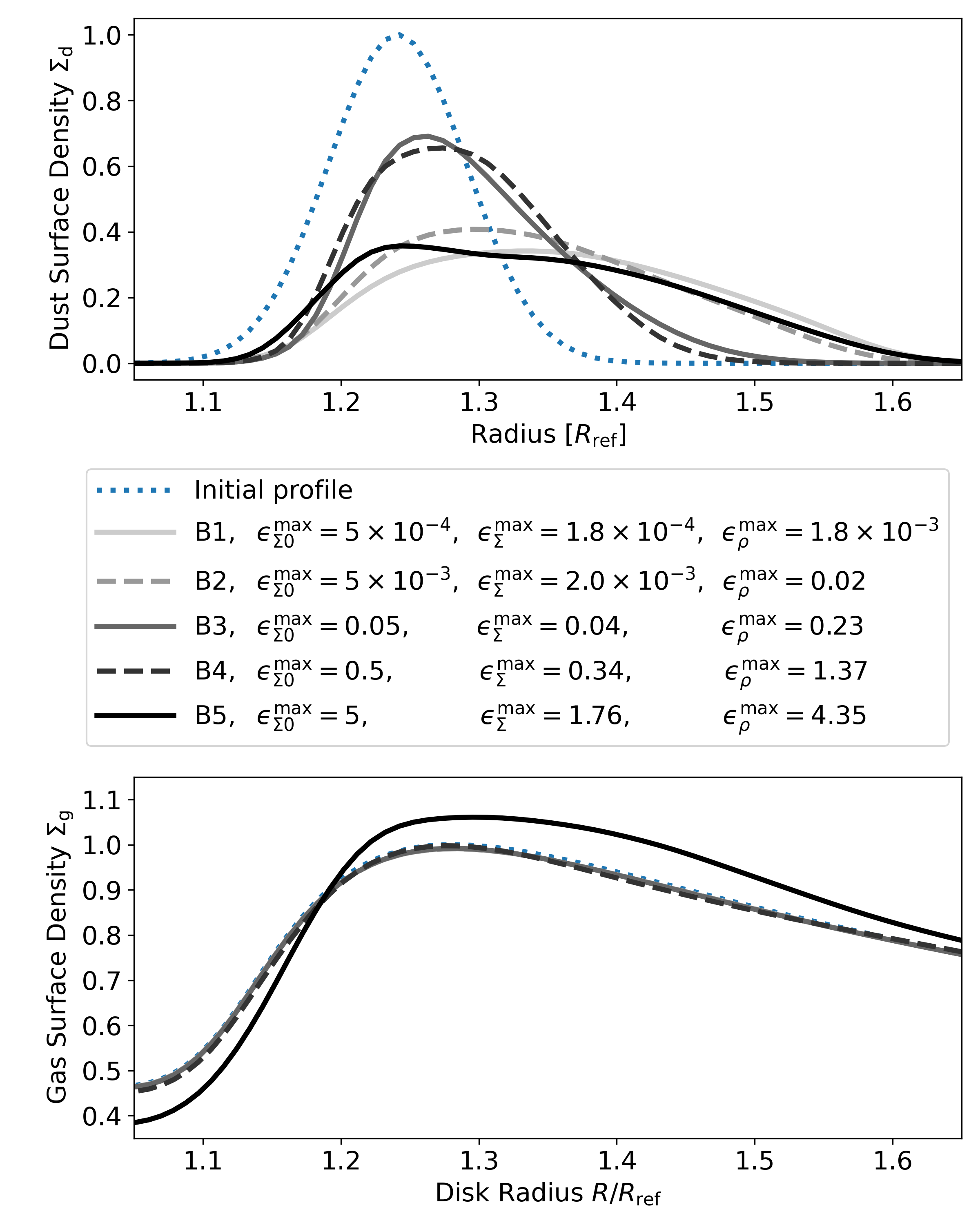}
\figcaption{
The azimuthally averaged dust surface density (\textit{top}) and gas surface density (\textit{bottom}) at $t = 1000T_{\rm ref}$ in type B models. All profiles are normalized to the corresponding maximum value in the initial profile. Since dust rings in different models are initialized with the same radial width but different dust load levels, we only compare the radial location and width of profiles in the top panel. The maximum values of the dust-to-gas surface density ratio $\epsilon_\Sigma^{\rm max}$, as well as its initial value $\epsilon_{\Sigma0}^{\rm max}$ and the corresponding midplane density ratio $\epsilon_\rho^{\rm max}$ are listed in the legends.
\label{fig:modelb}}
\end{figure}

To show that a gap-opening planet can widen dust rings on top of the effect of dust back-reaction, we make comparisons among type B models, which also include a Gaussian dust ring and a gas gap, but with different dust load levels. Different from type C models in Section \ref{sec:resultc}, the gas evolution and dust back-reaction are included in type B models. The top panel of Figure \ref{fig:modelb} shows how different levels of dust back-reaction would change the dust ring morphology on top of the same planet-related effects.

Firstly, we find that in Model B1 and B2, in which dust back-reaction are negligible due to low dust load, the dust rings are widened and moved outward as much as that in Model C1. Then in Model B3 and B4, in which the midplane dust-to-gas ratio is close to one, the rings become narrower compared with those in Model B1 and B2, and are closer to the pressure maximum.

The comparison between models of low and moderate dust load levels confirms that, dust back-reaction helps dust to resist effects that are related to the gas kinematics. When the dust back-reaction is negligible (Model B1 and B2) or even neglected (Model C1), the planetary wakes are not only capable of widening the dust ring, but also capable of providing an outward net mass flux together with the gas meridional flows, that carries well-coupled dust grains outward. When the dust back-reaction is moderately strong ($\epsilon_\rho \sim 1$), the dust ring can still be widened, but to a lesser extent; and it becomes harder to be carried outward by gas flows, acting like a damper to those effects. 

Therefore, we conclude that when $\epsilon_\rho$ increases from 0 to 1, the dust ring at the edge of the planet-opened gap would become narrower. It may seem contradicting to the result in \cite{kanagawa_impacts_2018}, which states effective dust back-reaction would flatten the global pressure profile and make the dust ring wider, but it is not. In Model B3 and B4, the maximum \textit{local} density ratio $\epsilon_\rho^{\rm max}$ is close to one, but the vertically integrated \textit{global} surface density ratio $\epsilon_\Sigma^{\rm max}$, which is the term monitored in the 2D disk model in \cite{kanagawa_impacts_2018}, is much smaller. This means even though the dust can deform the pressure profile close to the midplane via back-reaction, the gas on top of the midplane can compensate for that. Similar phenomena can be seen from the comparison between the axisymmetric (radial-vertical), unstratified disk models in \cite{taki_dust_2016} and the stratified models in \cite{onishi_planetesimal_2017}. In our models, this argument is supported by the bottom panel of Figure \ref{fig:modelb}, which shows that the gas density profiles from Model B1 to B4 are all consistent with each other. When it comes to both $\epsilon_\Sigma^{\rm max} > 1$ and $\epsilon_\rho^{\rm max} > 1$ in Model B5, the dust ring becomes so dust-rich that it is capable of deforming the global gas density profile via overwhelming back-reaction. Then the dust ring becomes wider as a result of the pressure profile being flattened, agreeing with \cite{kanagawa_impacts_2018}.

To conclude, moderate dust back-reaction ($\epsilon_\rho \sim 1$) tends to make the dust ring narrower by damping the planetary perturbations. However, when $\epsilon_\Sigma > 1$ and $\epsilon_\rho > 1$, the ring expands itself during the process of deforming the global pressure profile via overwhelming back-reaction, with the widening effect being no more attributable to the planet-related effects. We note that Model B3 provides the most similar dust load level to our type A models ($\epsilon_\rho^{\rm max} \sim 0.5$) in Section \ref{sec:resulta} and conventional thoughts of protoplanetary disks. Therefore, the dust rings in our type A models tend to demonstrate the minimum width under the net effect of planetary perturbations and dust back-reaction.

\subsection{How Much Can the Dust Ring Be Widened by a Planet?} \label{sec:resulta}

The radial FWHM of an equilibrated dust ring $w_{\rm d}$ is determined by the balance between the concentration effect due to pressure gradient and the expansion effect due to diffusion. Therefore, in our models without the turbulence-induced dust diffusion, if the planet can widen dust rings and maintain this effect, the dust rings can be modelled as being widened by an effective diffusion with a diffusion tensor $\mathcal{D}$. In this section, we first try to establish a quantification of $\mathcal{D}$ via gradient diffusion hypothesis, and then derive the relation between $\mathcal{D}$ and $w_{\rm d}$.

\subsubsection{Quantify $\mathcal{D}$ with Gradient Diffusion Hypothesis} \label{sec:original_eq}

The planet-related effects expand dust rings via introducing fluctuations to the disk, which may be modeled as diffusion. If so, we would like to quantify those fluctuations in order to obtain the diffusion coefficients. In our disk models, a physical variable \textit{A} can be azimuthally decomposed to $A = \langle A\rangle + \Delta A$, where $\langle A\rangle$ is the mean field and $\Delta A$ is the fluctuation term with $\langle\Delta A\rangle = 0$. Applying the decomposition to Equation \ref{eq:dustcont} then taking the azimuthal average, we get
\begin{equation}
    \label{eq:prhodpt}
    \frac{\partial \langle\rho_{\rm d}\rangle}{\partial t} =
    -\nabla\cdot(\langle\rho_{\rm d}\rangle\langle\W\rangle)
    -\nabla\cdot\langle\Delta\rho_{\rm d}\Delta\W\rangle.
\end{equation}

The product term of fluctuations above is associated with diffusion via the gradient diffusion hypothesis (e.g., \citealt{cuzzi_particle-gas_1993, tominaga_revised_2019})
\begin{equation} \label{eq:grad-diff}
    \langle\Delta\rho_{\rm d}\Delta\W\rangle = -\mathcal{D}\nabla\langle\rho_{\rm d}\rangle,
\end{equation}
where
\begin{equation}
    \mathcal{D} = 
    \begin{bmatrix}
        \mathcal{D}_{RR} & \mathcal{D}_{RZ} \\
        \mathcal{D}_{ZR} & \mathcal{D}_{ZZ}
    \end{bmatrix}
\end{equation}
is a diffusion tensor that describes the dust diffusion in the frame of the disk. We note that only diffusion in $\{R,Z\}$ directions are considered here, as the $\phi$ components become irrelevant in Equation \ref{eq:prhodpt}. Equation \ref{eq:grad-diff} provides a quantification of $\mathcal{D}$, but it does not give a unique solution because it only has two equations for the four components of $\mathcal{D}$. Therefore, we assume $\mathcal{D}_{RZ} = \mathcal{D}_{ZR} = 0$, indicating radial gradient of density does not contribute to vertical diffusion, and vice versa, to reduce number of unknowns. Then for conciseness, we use $\mathcal{D}_R$ and $\mathcal{D}_Z$ to denote $\mathcal{D}_{RR}$ and $\mathcal{D}_{ZZ}$, respectively. 
Nevertheless, $\mathcal{D}$ at the dust peak (where $\nabla \langle\rho_{\rm d}\rangle = 0$), which is the key for the widening effect of the dust ring, cannot be quantified via Equation \ref{eq:grad-diff}. Therefore, we quantify $\mathcal{D}$ via balancing the advection term and the diffusion term in Equation \ref{eq:prhodpt}, in a steady state where $\partial \langle\rho_{\rm d}\rangle/\partial t = 0$. Later in Section \ref{sec:d_to_w} we will show that, a profile of $\mathcal{D}$ with $\nabla\mathcal{D} \sim 0$ in the extent of the dust ring, and a scenario where advection and diffusion are balanced in individual directions, are preferred while associating $\mathcal{D}$ with $w_{\rm d}$. Therefore, here we would expect a constant $\overline{\mathcal{D}}$ at the dust ring with
\begin{align} 
    \label{eq:balancez}
    \nabla_Z\cdot(\langle\rho_{\rm d}\rangle\langle W_Z\rangle) &= 
    \overline{\mathcal{D}}_Z\nabla_Z^2\langle\rho_{\rm d}\rangle \\
    \label{eq:balancer}
    \nabla_R\cdot(\langle\rho_{\rm d}\rangle\langle W_R\rangle) &= 
    \overline{\mathcal{D}}_R\nabla_R^2\langle\rho_{\rm d}\rangle.
\end{align}

We then run type A models to validate this method. Different from those in type B and C models, gas and dust densities in type A models are both initialized to power-law radial profiles, with no gaps or rings. When simulations begin, the planet mass increases from zero and gradually opens a gap, which eventually leads to the formation of a dust ring. Type A models are differentiated by the planet mass and Stokes number, with the model of $\{M_{\rm p}, {\rm St}_{\rm 0,ref}\} = \{3\times10^{-4}M_\star, 10^{-3}\}$ being the fiducial one. To provide near-steady states, all type A models are run for $3000T_{\rm ref}$. 

\begin{figure*}[tp] 
\includegraphics[width = \textwidth]{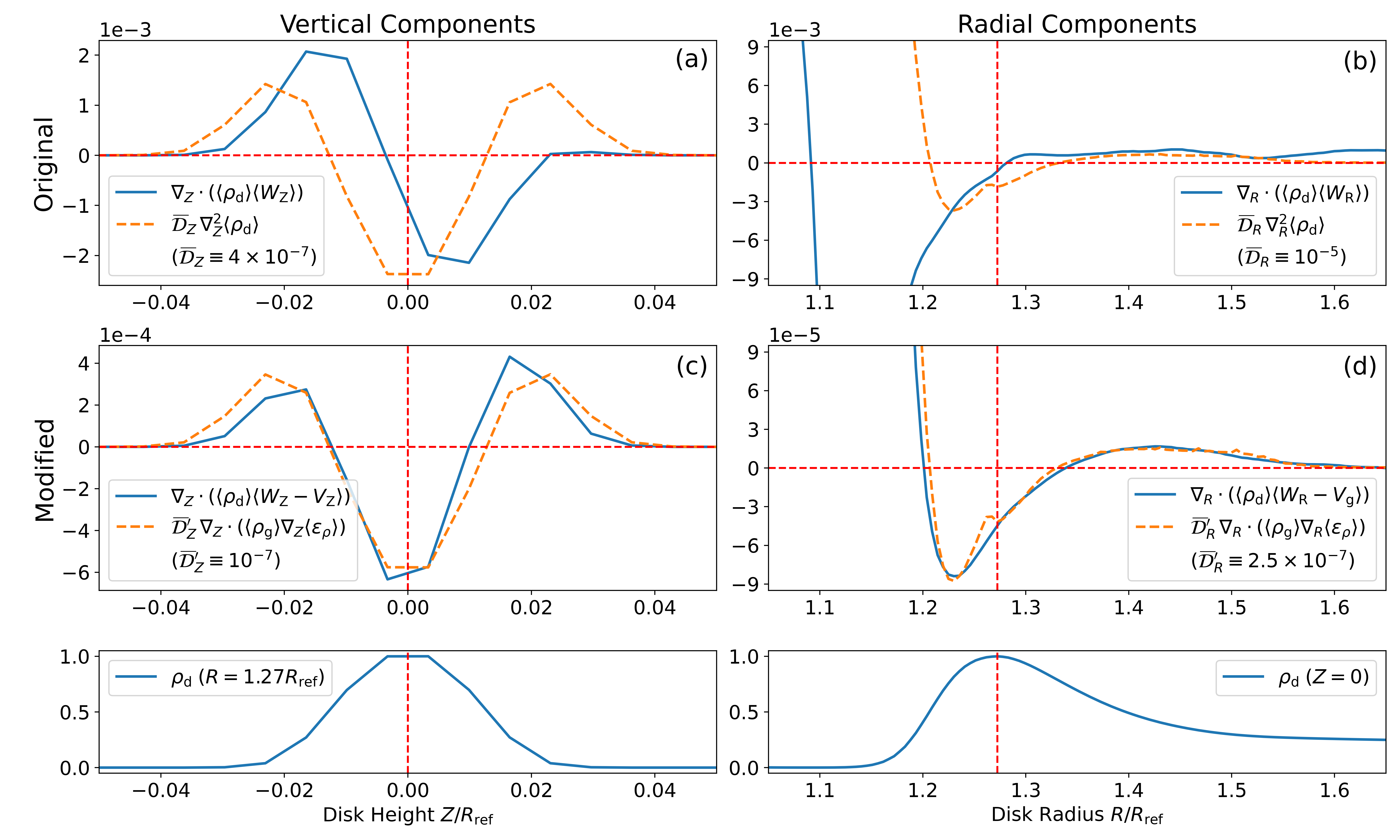}
\figcaption{
($a$): The vertical profile of terms in Equation \ref{eq:balancez} at the dust peak. 
($b$): The radial profile of terms in Equation \ref{eq:balancer} in the midplane. 
($c$): The vertical profile of vertical components in Equation \ref{eq:adv-diff} at the dust peak. 
($d$): The radial profile of radial components in Equation \ref{eq:adv-diff} in the midplane. 
All profiles are evaluated in the fiducial type A model at $t = 3000T_{\rm ref}$. 
The demonstrative constant diffusion coefficients $\overline{\mathcal{D}}$ and $\overline{\mathcal{D}}^\prime$, in the unit of $R_{\rm ref}^2 \Omega_{\rm K,ref}$, are chosen such that the two curves in each panel are at the same order of magnitude.
Panel $a$ and $b$ represent the original gradient diffusion hypothesis in Section \ref{sec:original_eq}.
Panel $c$ and $d$ represent the modified gradient diffusion hypothesis in Section \ref{sec:modified_eq}.
Panel $a$ and $c$ are normalized to the midplane dust density at the dust peak.
Panel $b$ and $d$ are normalized to the radial profile of the midplane dust density.
The vertical red dashed lines in the left column denote $Z = 0$, the ones in the right column denote the dust peak at $R = 1.27R_{\rm ref}$.
The horizontal red dashed lines denote $y = 0$. 
The normalized vertical and radial profiles of dust density are shown in the bottom for reference.
\label{fig:compare}}
\end{figure*}

The profiles of terms in Equation \ref{eq:balancez} and \ref{eq:balancer}, evaluated vertically and radially across the dust peak, are shown in Figure \ref{fig:compare} (panel $a$--$b$). We choose constant values of $\overline{\mathcal{D}}$ and find no good match between the advection term and the diffusion term. Panel $a$ shows that, while the vertical profile of the diffusion term is an even function relative to the midplane, the profile of the advection term is close to an odd one. In our model, this qualitative mismatch is resulted from a meridional flow that does not only cross the midplane but also penetrates through the dust layer, which breaks the symmetry along the midplane that is conventionally assumed in previous studies. The meridional flows affect the radial profiles as well. In panel $b$, although the two profiles are similar in a large radial extent, they have different zero-crossings, indicating negative diffusion coefficient in part of the ring. Besides, the two profiles diverge at $R \sim 1.23R_{\rm ref}$, which is still within the half width at half maximum to the peak. In addition, we note that the two mismatches in the radial and vertical direction do not compensate for each other, and they cannot be fixed by setting non-zero values of $\overline{\mathcal{D}}_{RZ}$ and $\overline{\mathcal{D}}_{ZR}$ (see Appendix \ref{app:mismatch}). Therefore, we conclude that this method of quantifying a constant $\overline{\mathcal{D}}$ at the dust ring using the gradient diffusion hypothesis is not applicable to our disk models with active meridional flows induced by gap-opening planets.

\subsubsection{Modifying the Gradient Diffusion Hypothesis} \label{sec:modified_eq}

In our models, the advection of well-coupled dust is correlated to the gas kinematics. However, the method in the above section does not have any explicit dependence on the gas. Here we show that, after taking the non-trivial gas density and velocity field into consideration, a \textit{modified} gradient diffusion hypothesis can model fluctuations of the relative motion between gas and dust. 

To address the effects of bulk motion and fluctuation of gas, we first apply the azimuthal decomposition to Equation \ref{eq:gascont} to get 
\begin{equation}
    \label{eq:prhogpt}
    \frac{\partial \langle\rho_{\rm g}\rangle}{\partial t} =
    -\nabla\cdot(\langle\rho_{\rm g}\rangle\langle\V\rangle)
    -\nabla\cdot\langle\Delta\rho_{\rm g}\Delta\V\rangle.
\end{equation}
Considering both fluctuation terms in Equation \ref{eq:prhodpt} and \ref{eq:prhogpt}, we modify Equation \ref{eq:grad-diff} (to be justified a posteriori) to 
\begin{align}
    \nonumber 
    \nabla\cdot\langle\Delta\rho_{\rm d}\Delta\W\rangle 
      -\langle\epsilon_\rho\rangle&\nabla\cdot\langle\Delta\rho_{\rm g}\Delta\V\rangle \\ \label{eq:grad-diff-mod}
    = -&\nabla\cdot\bigg(\mathcal{D}^\prime\langle\rho_{\rm g}\rangle\nabla\langle\epsilon_\rho\rangle\bigg),
\end{align}
where $\langle\epsilon_\rho\rangle = \langle\rho_{\rm d}\rangle/\langle\rho_{\rm g}\rangle$ is the mean field dust-to-gas ratio. Here, $\mathcal{D}^\prime$ is the diffusion tensor that describes the diffusion of dust \textit{relative to} the gas, also with non-diagonal elements assumed to be zero. Instead of the steady state for both gas and dust with $\partial \langle\rho_{\rm d}\rangle/\partial t = \partial \langle\rho_{\rm g}\rangle/\partial t = 0$, we suggest that $\mathcal{D}^\prime$ can be quantified in a more general case with
\begin{equation} \label{eq:lagrangian}
    \frac{{\rm D}\langle\epsilon_\rho\rangle}{{\rm D}t} 
    = \frac{\partial\langle\epsilon_\rho\rangle}{\partial t} 
    + \langle\V\rangle\cdot\nabla\langle\epsilon_\rho\rangle = 0.
\end{equation}
Equation \ref{eq:lagrangian} is an advection equation that says the Lagrangian derivative of $\langle\epsilon_\rho\rangle$ is zero. That is, the dust-to-gas ratio following a gas parcel remains constant. Considering Equation \ref{eq:prhodpt}, \ref{eq:prhogpt}, and \ref{eq:grad-diff-mod}, Equation \ref{eq:lagrangian} is equivalent to (see Appendix \ref{app:derivation})
\begin{equation} \label{eq:adv-diff}
    \nabla\cdot\bigg(\langle\rho_{\rm d}\rangle\langle\W-\V\rangle\bigg) +
    \nabla\cdot\bigg(\mathcal{D}^\prime\langle\rho_{\rm g}\rangle\nabla\langle\epsilon_\rho\rangle\bigg) = 0,
\end{equation}
which says the relative advection between gas and dust balances the relative diffusion between them. 

We note that the dust and gas in the above state are not necessarily steady ($\partial/\partial t \neq 0$). For example, Equation \ref{eq:lagrangian} and \ref{eq:adv-diff} are still applicable when the dust and gas are moving together in space due to bulk motions while maintaining no relative evolution. In other words, for the purpose of associating $\mathcal{D}^\prime$ with $w_{\rm d}$ in Section \ref{sec:d_to_w}, we only need an equilibrated $w_{\rm d}$. We also note that the combination of Equation \ref{eq:lagrangian} and \ref{eq:adv-diff} agrees with the correct\footnote{Other forms may lead to inappropriate derivations with wrong coefficients or unphysical terms} advection-diffusion equation in the context of protoplanetary disks discussed in \cite{desch_formulas_2017}.

Figure \ref{fig:compare} (panel $c$--$d$) shows the profiles of terms in Equation \ref{eq:adv-diff}. We find that the relative diffusion and advection balance in both directions, and the match of profiles can be obtained by constant  $\overline{\mathcal{D}}_R^\prime$ and $\overline{\mathcal{D}}_Z^\prime$, but with different values. Therefore, we conclude that the radial component of diffusion coefficient $\mathcal{D}_R^\prime$ at the dust peak can be quantified using the modified gradient diffusion hypothesis, and it can be approximated as a constant value in the radial extent of the dust ring.

\subsubsection{Associating $\mathcal{D}^\prime$ with the Dust Ring Width $w_{\rm d}$} \label{sec:d_to_w}

After quantifying $\mathcal{D}_R^\prime$, the dust ring width $w_{\rm d}$ can be obtained with certain dust and gas density profiles. For simplicity, in this section we focus on the correlation between $\mathcal{D}_R^\prime$ and $w_{\rm d}$ within the immediate vicinity of the dust peak in the midplane. And for conciseness, all variables in this section are azimuthally averaged by default. Although $\rho_{\rm d}$ and $\rho_{\rm g}$ may not peak at the same location, we assume $\partial \rho_{\rm g}/\partial R \sim 0$, and consequently $\partial \epsilon/\partial R \sim 0$, in the vicinity\footnote{This approximation is appropriate in our type A models with moderate ($\epsilon_\rho \sim 1$) dust back-reaction.}. 

Since in Figure \ref{fig:compare} we show that the balance in Equation \ref{eq:adv-diff} can be reached in individual directions, here we write the radial part of it:
\begin{equation} \label{eq:2nd_order}
    - \frac{\partial}{\partial R}\bigg[R\rho_{\rm d}(W_R - V_R)\bigg]
    + \frac{\partial}{\partial R}\bigg(R\mathcal{D}_R^\prime\rho_{\rm g}\frac{\partial \epsilon_\rho}{\partial R}\bigg)
    = 0.
\end{equation}
With the dust kinematics in our models agreeing with the terminal velocity approximation (Equation \ref{eq:approx}), after rewriting $\rho_{\rm g}\partial\epsilon_\rho/\partial R = \partial\rho_{\rm d}/\partial R - \epsilon_\rho\partial\rho_{\rm g}/\partial R$, and considering the locally isothermal equation of state and the power-law radial temperature profile, we get
\begin{align} \label{eq:1st_order}
    \frac{\partial}{\partial R}\bigg[\epsilon_\rho R\bigg(\frac{\tau_{\rm s}c_{\rm s}^2}{1+\epsilon_\rho} + \mathcal{D}_R^\prime\bigg)\frac{\partial\rho_{\rm g}}{\partial R}& \nonumber \\
    -\epsilon_\rho qR\frac{\tau_{\rm s}c_{\rm s}^2}{1+\epsilon_\rho}\frac{\rho_{\rm g}}{R}&-R\mathcal{D}_R^\prime\frac{\partial\rho_{\rm d}}{\partial R}\bigg] = 0.
\end{align}
Assuming $\partial \mathcal{D}_R^\prime/\partial R \sim 0$ in the vicinity\footnote{This approximation is discussed in Section \ref{sec:modified_eq}.}, and recalling previous assumptions of $\partial \rho_{\rm d}/\partial R \sim \partial \rho_{\rm g}/\partial R \sim \partial \epsilon_\rho/\partial R \sim 0$ there, the terms with first-order derivatives in Equation \ref{eq:1st_order} can be dropped, and Equation \ref{eq:1st_order} can be approximated to
\begin{equation} \label{eq:0th_order}
    \epsilon_\rho\bigg(\frac{\tau_{\rm s}c_{\rm s}^2}{1+\epsilon_\rho} + \mathcal{D}_R^\prime\bigg)\frac{\partial^2\rho_{\rm g}}{\partial R^2}+\frac{q^2}{2}\frac{\tau_{\rm s}c_{\rm s}^2}{1+\epsilon_\rho}\frac{\rho_{\rm d}}{R}-\mathcal{D}_R^\prime\frac{\partial^2\rho_{\rm d}}{\partial R^2} = 0.
\end{equation}
To associate Equation \ref{eq:0th_order} with the radial width of disk structures, we assume both gas and dust density profiles in the vicinity are Gaussian:
\begin{align}
    \rho_{\rm g}(R) &= \rho_{\rm g}(R_{\rm g})\times \exp\bigg[-\frac{(R-R_{\rm g})^2}{2w_{\rm g}^2}\bigg] \\
    \rho_{\rm d}(R) &= \rho_{\rm d}(R_{\rm d})\times \exp\bigg[-\frac{(R-R_{\rm d})^2}{2w_{\rm d}^2}\bigg].
\end{align}
$R_{\rm g}$ and $R_{\rm d}$ are the radii where the gas and dust density profile peak, and $w_{\rm g}$ and $w_{\rm d}$ are the widths of the dust ring and the gas bump. By approximating $w_{\rm g}^2 \gg (R-R_{\rm g})^2$ within the vicinity of the dust peak, there are
\begin{align}
    \frac{\partial^2 \rho_{\rm g}}{\partial R^2} 
        &= \rho_{\rm g}\frac{(R-R_{\rm g})^2 - w_{\rm g}^2}{w_{\rm g}^4} \approx -\frac{\rho_{\rm g}}{w_{\rm g}^2} \\ 
    \frac{\partial^2 \rho_{\rm d}}{\partial R^2} 
        &= \rho_{\rm d}\frac{(R-R_{\rm d})^2 - w_{\rm d}^2}{w_{\rm d}^4} \approx -\frac{\rho_{\rm d}}{w_{\rm d}^2}.
\end{align}
Then taking those back to Equation \ref{eq:0th_order}, we finally get
\begin{equation} \label{eq:wd}
    w_{\rm d} = \sqrt{\dfrac{\mathcal{D}_R^\prime}{\bigg(\dfrac{\tau_{\rm s}c_{\rm s}^2}{1+\epsilon_\rho} + \mathcal{D}_R^\prime\bigg)\dfrac{1}{w_{\rm g}^2} - \dfrac{\tau_{\rm s}c_{\rm s}^2}{1+\epsilon_\rho}\dfrac{q^2}{2R^2}}}, 
\end{equation}
which shows how much the dust ring can be widened by the diffusion-like behavior, given the radial location, the stopping time, the dust-to-gas ratio, the sound speed profile, and the gas structure at the dust ring. We note that in the limit of $w_{\rm g}^2 \ll 2R^2/q^2$ and $(1+\epsilon_\rho)\mathcal{D}_R^\prime \ll \tau_{\rm s}c_{\rm s}^2$, Equation \ref{eq:wd} can be approximated to 
\begin{equation} \label{eq:wd/wg}
    w_{\rm d} = w_{\rm g}\sqrt{\dfrac{(1+\epsilon_\rho)\mathcal{D}_R^\prime}{\tau_{\rm s}c_{\rm s}^2}},
\end{equation}
which is similar to the Equation 46 in \cite{dullemond_disk_2018}, but with different definitions of parameters.

%% file: Section_4.tex
\section{Discussion} \label{sec:discussion}
\subsubsection*{Connection to the Turbulent Viscosity}

In this paper, we provide a way to associate the planet-related diffusion coefficient $\mathcal{D}_R^\prime$ with the dust ring width $w_{\rm d}$. However, in disk observations, $\mathcal{D}_R^\prime$ is neither a measurable nor a property that can be easily constrained by measurables, making it hard to infer any properties of the suspected planet using the dust ring width. Here we discuss the feasibility of associating $\mathcal{D}_R^\prime$ of the dust component with the Reynolds stress $\mathcal{R}$ of the gas component, which describes the radial turbulent angular momentum transport in the disk, and is more easily constrained in observations.

In our 3D disk model, we calculate the azimuthally averaged profile of the Reynolds stress via
\begin{equation}
    \mathcal{R} = \langle\rho_{\rm g}\Delta V_{\rm R}\Delta V_\phi\rangle,
\end{equation}
where $\Delta V$ is the fluctuation of gas velocity to its mean field. Then the turbulent viscosity parameter $\alpha_{\rm turb}$ \citep{shakura_reprint_1973} can be obtained via
\begin{equation} \label{eq:alpha}
    \alpha_{\rm turb} = \mathcal{R}/\langle P\rangle = \frac{\nu_{\rm turb}}{c_{\rm s}H},
\end{equation}
where $\nu_{\rm turb}$ is the measured kinematics viscosity and may not be identical to the implemented $\nu$ in the model. Since the ratio between the momentum diffusivity (i.e., the kinematics viscosity) and the mass diffusivity (i.e., the diffusion coefficient) is the Schmidt number Sc, Equation \ref{eq:wd} and \ref{eq:wd/wg} can be rewritten to
\begin{equation} \label{eq:wd_psi_complicated}
    w_{\rm d} = \left[\left(1 + \frac{1}{\psi^2}\right)\frac{1}{w_{\rm g}^2}-\frac{1}{\psi^2}\frac{q^2}{2R^2}\right]^{-1/2}
\end{equation}
and
\begin{equation} \label{eq:wd_psi_simplified}
    w_{\rm d} = w_{\rm g}\psi,
\end{equation}
where 
\begin{equation}
    \psi = \sqrt{\frac{(1+\epsilon_\rho)\mathcal{D}_R^\prime}{\tau_{\rm s}c_{\rm s}^2}} = \sqrt{\frac{(1+\epsilon_\rho)\alpha_{\rm turb}}{{\rm St}\,{\rm Sc}}}.
\end{equation}

\begin{figure}[tp] 
\includegraphics[width = 0.47\textwidth]{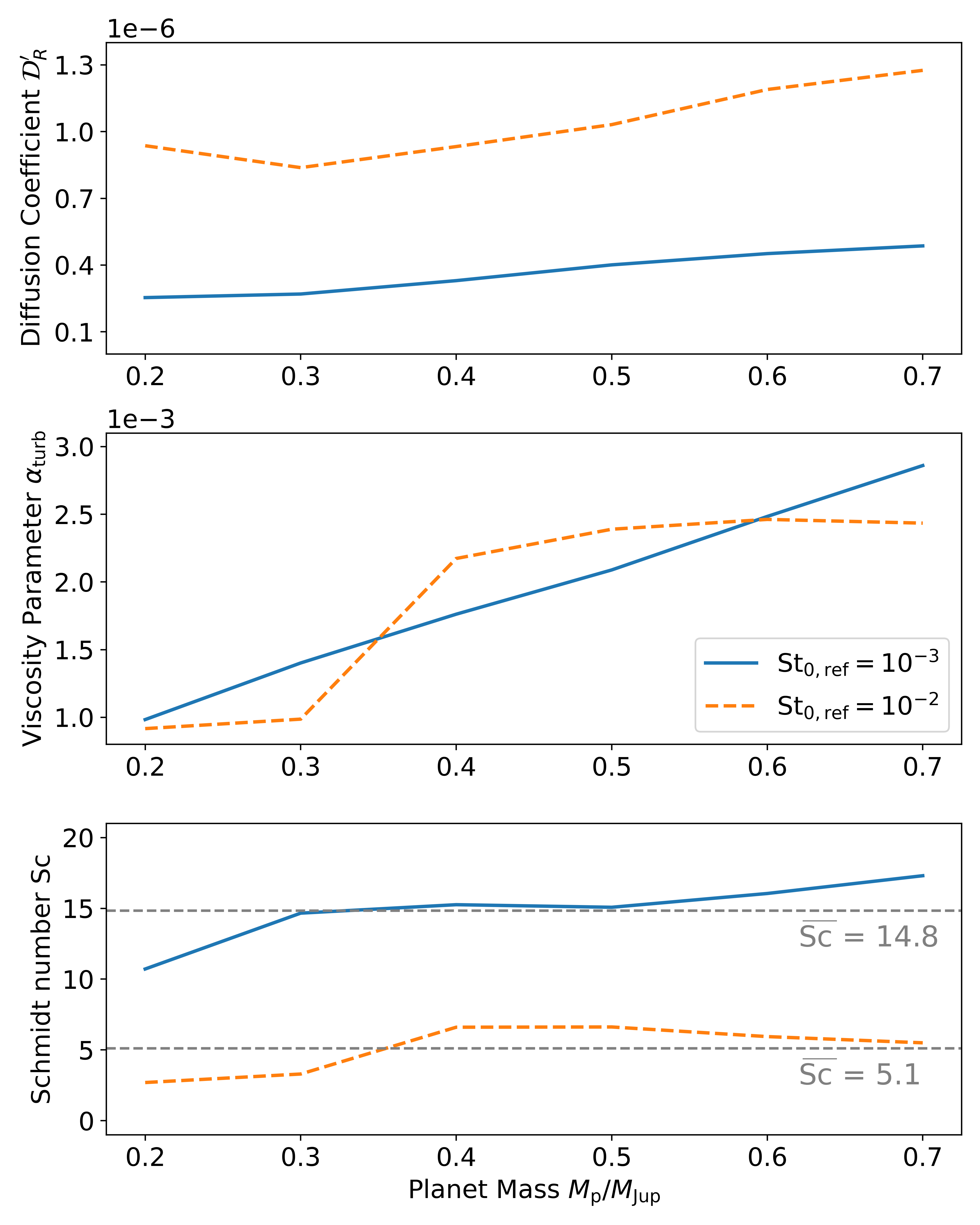}
\figcaption{The radial diffusion coefficient $\mathcal{D}_R^\prime$ (\textit{top}) from Equation \ref{eq:2nd_order}, the turbulent viscosity parameter $\alpha_{\rm turb}$ (\textit{middle}) from Equation \ref{eq:alpha}, and the corresponding Schmidt number Sc (\textit{bottom}) at the dust peak in different type A models. All profiles are evaluated in the midplane at $t = 3000T_{\rm ref}$. The unit of $\mathcal{D}_R^\prime$ in the top panel is $R_{\rm ref}^2 \Omega_{\rm K,ref}$. Horizontal dashed lines in the bottom panel mark the averaged Schmidt number $\overline{\rm Sc}$ for models with the same ${\rm St}_{\rm 0,ref}$.
\label{fig:viscparam}}
\end{figure}

Figure \ref{fig:viscparam} shows the measured $\mathcal{D}_R^\prime$, $\alpha_{\rm turb}$, and the corresponding Schmidt number at the dust peak in type A models with different planet masses and Stokes numbers of dust. Our resulted $\alpha_{\rm turb}$ values agree with the accretion levels estimated in disks with one or more planets of a few Earth masses in previous works \citep{goodman_planetary_2001, fung_save_2017}. In the meantime, we show that the resulted Schmidt number is larger than the conventionally assumed of order unity. These large Sc numbers suggest that the planet-related transport of mass and momentum may depend on the specific form of turbulence or perturbation. Therefore, we caution about the assumption of ${\rm Sc} = 1$ in studies on turbulent effects in protoplanetary disks. For models with the same ${\rm St}_{\rm 0,ref}$, we find the Schmidt numbers are similar. Whether this is a coincidence, why Sc decreases with increasing St, and how Sc changes with other disk parameters, will be investigated in the future. 
 
Overall, it provides a possible avenue to estimate the property of the suspected planet from disk observations. Since Sc is only sensitive to St in our setting (bottom panel of Figure \ref{fig:viscparam}), measurements of the size and St of dust at the ring via multi-wavelength dust observations and spectral energy distribution modeling \citep[e.g.,][]{guidi_distribution_2022} or mm-wavelength polarization observations \citep[e.g.,][]{kataoka_grain_2016} may lead to constraints on Sc. Since the $\alpha_{\rm turb}$ value and the dust ring width $w_{\rm d}$ may be obtained from gas and dust observations \citep[e.g.,][]{flaherty_measuring_2020}, the gas bump width $w_{\rm g}$ may be constrained from Equation \ref{eq:wd_psi_simplified} with certain assumptions on the level of $\epsilon_\rho$. Finally, since the radial gas density profile modified by a gap-opening planet, which includes the gas bump, can be estimated analytically \citep[e.g.,][]{duffell_simple_2015, duffell_empirically_2020}, the orbital radius and mass of the suspected planet may be obtained. 

%% file: Section_5.tex
\section{Conclusion} \label{sec:conclusion}

In this paper, we use 3D hydrodynamic simulations to study the dust kinematics in protoplanetary disks where a planet is present. Our main findings are:

\begin{enumerate}
    \item Compared with dust rings trapped at axisymmetric pressure bumps, those trapped at planet-induced pressure bumps featuring density waves are widened by planet-disk interactions. For dust rings composed of small (${\rm St} \lesssim 10^{-2}$) grains, the fluctuations in the gas velocity field due to planetary wakes are most responsible for the widening effect.
    \item Moderate dust back-reaction with local dust-to-gas ratio $\lesssim 1$ tends to narrow dust rings under the planet-related effects, compared with the cases where dust back-reaction is negligible. However, overwhelming dust back-reaction with dust-to-gas ratios of both volumetric and surface density $>1$ would lead to the dust ring expanding itself while deforming the global pressure profile. In the overwhelmingly high dust mass regime, both dust back-reaction and planet-related effects are in effect, but the former takes dominance.
    \item The widening effect of dust rings due to planet-related effects can be modelled by our modified gradient diffusion hypothesis, and can be quantified by a diffusion coefficient on the order of $10^{-7}\text{--}10^{-6} R^2 \Omega_{\rm K}$. We note that the conventional gradient diffusion hypothesis with globally constant diffusion coefficients is not applicable to our 3D disk models with planet-induced meridional flows.
    \item We show that the widening effect can also be quantified by the Reynolds stress, with the corresponding turbulent viscous parameter $\alpha_{\rm turb}$ on the order of $10^{-3}$. However, we caution about the Schmidt number being greater than order unity. It suggests that a high momentum diffusivity of gas does not always translate to high mass diffusivity of dust, even for well-coupled dust ($\tau_{\rm s} \ll \Omega_{\rm K}^{-1}$) in the disk.
\end{enumerate}

We thank the anonymous referee for the helpful suggestions that improved the quality of the paper. We thank He-Feng Hsieh and Jeffrey Fung for useful discussions. We also thank Ya-Wen Tang for sharing the processed ALMA data of AB Aur with us. Simulations were carried out on the TAIWANIA-2 GPU cluster hosted by the National Center for High-Performance Computing. M.-K.L. is supported by the National Science and Technology Council (grants 107-2112-M-001-043-MY3, 110-2112-M-001-034-, 111-2112-M-001-062-, 110-2124-M-002-012-, 111-2124-M-002-013-) and an Academia Sinica Career Development Award (AS-CDA-110-M06). J.B. and R.D. are supported by the Natural Sciences and Engineering Research Council of Canada. R.D. acknowledges support from the Alfred P. Sloan Foundation via a Sloan Research Fellowship.

\appendix
\section{Disk Plots of Model C1 and C5} \label{app:eclip}

\begin{figure}
\includegraphics[width = \textwidth]{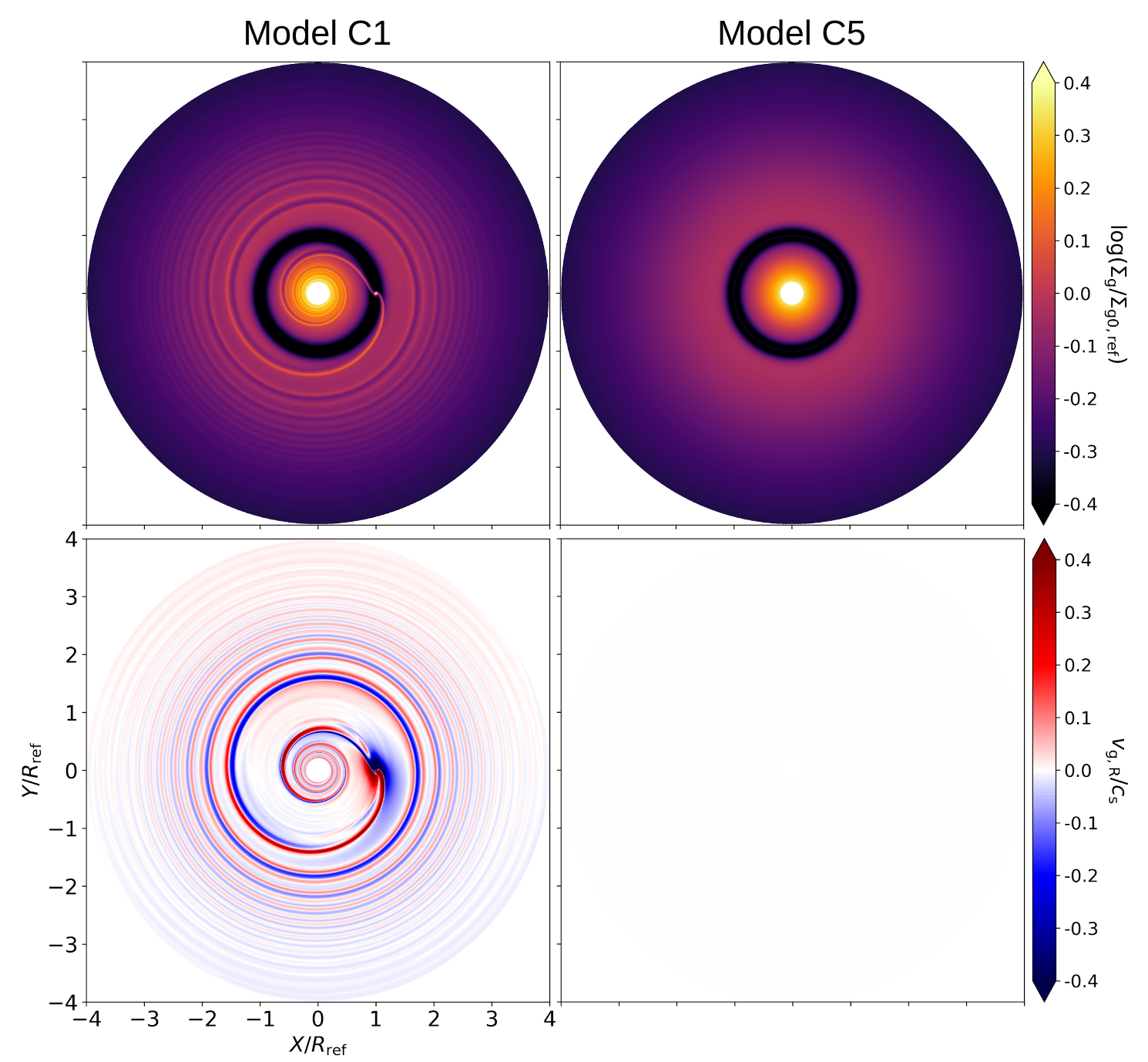}
\figcaption{
The gas surface density (\textit{top panels}) and midplane gas radial velocity (\textit{bottom panels}) of Model C1 (\textit{left panels}) and C5 (\textit{right panels}). Since gas evolution is stalled in type C models, all four panels are time-invariant. In Model C1, planet-related spirals in both gas density and velocity fields are preserved, whereas those in Model C5 are erased by azimuthal averaging. The bottom right panel is all white because the midplane gas in Model C5 has zero radial velocity.
\label{fig:eclip}}
\end{figure}

To show the differences in the gas initialization between Model C1 and C5, we plot the face-on view of gas surface density and midplane gas radial velocity in Figure \ref{fig:eclip}.

\section{The Mismatch Between Advection and Diffusion With the Gradient Diffusion Hypothesis} \label{app:mismatch}

\begin{figure}
\centering
\includegraphics[width = 0.75\textwidth]{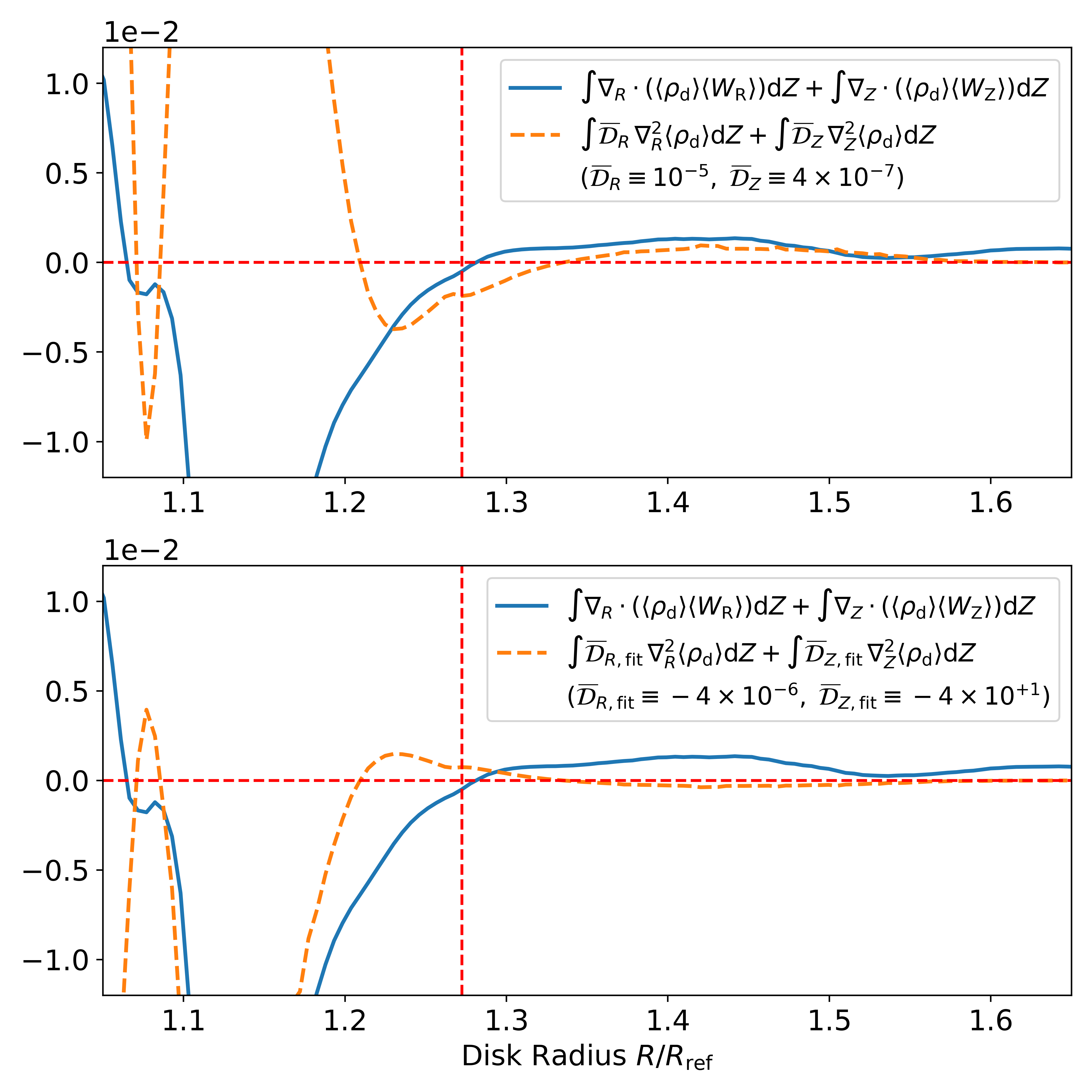}
\figcaption{The radial profile of vertically integrated advection and diffusion terms with the gradient diffusion hypothesis in Section \ref{sec:original_eq}. All profiles are normalized to the dust surface density profile and are evaluated in the fiducial type A model at $t = 3000T_{\rm ref}$. $\overline{\mathcal{D}}_R$ and $\overline{\mathcal{D}}_Z$ in the top panel are identical to the ones in Figure \ref{fig:compare} ($a$--$b$). In the bottom panel, $\overline{\mathcal{D}}_R$ and $\overline{\mathcal{D}}_Z$ are fitted within the plotted range for the best match. The unit of $\overline{\mathcal{D}}_R$ and $\overline{\mathcal{D}}_Z$ is $R_{\rm ref}^2 \Omega_{\rm K,ref}$. The horizontal red dashed line denotes $y = 0$. The vertical red dashed line denotes the dust peak at $R = 1.27R_{\rm ref}$.
\label{fig:compare1}}
\end{figure}

Figure \ref{fig:compare1} shows the vertically integrated advection and diffusion terms from Equation \ref{eq:prhodpt} and \ref{eq:grad-diff}. In the top panel, the constant diffusion tensor elements $\overline{\mathcal{D}}_R$ and $\overline{\mathcal{D}}_Z$ are identical to the ones in the top row of Figure \ref{fig:compare}. We find that the two profiles still do not match, similar to those in Figure \ref{fig:compare} ($a$--$b$), even when both radial and vertical components are taken into consideration. In the bottom panel, $\overline{\mathcal{D}}_R$ and $\overline{\mathcal{D}}_Z$ are numerically fitted within the plotted range for the best match between the two profiles. While we do not find a match as good as the ones in the middle row of Figure \ref{fig:compare} with the modified gradient diffusion hypothesis, the fitted $\overline{\mathcal{D}}_R$ and $\overline{\mathcal{D}}_Z$ are also not realistic. 

\begin{figure}
\centering
\includegraphics[width = 0.75\textwidth]{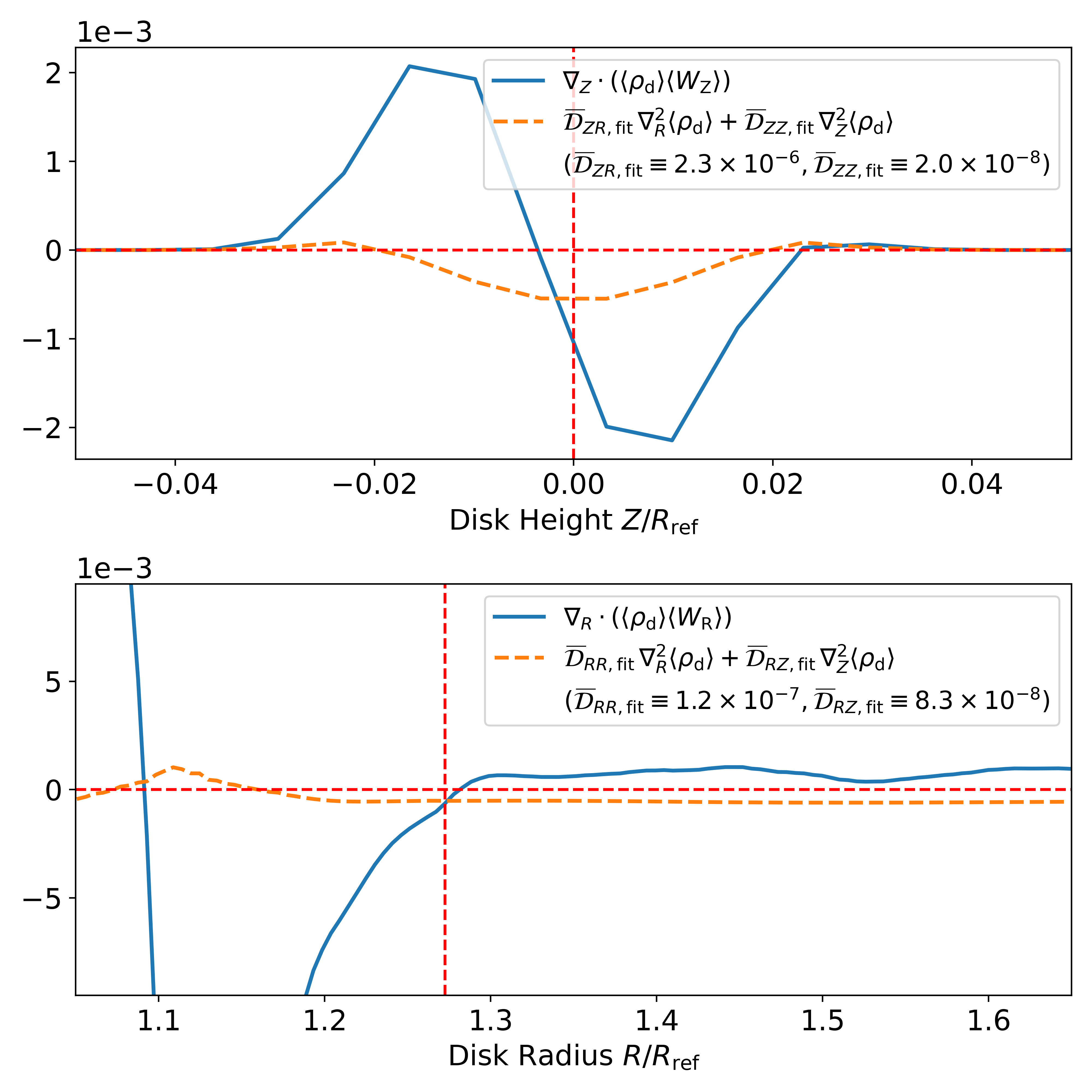}
\figcaption{Similar to the panel $a$ and $b$ in Figure \ref{fig:compare}, but allowing $\mathcal{D}_{ZR}$ and $\mathcal{D}_{RZ}$ to be non-zero. All four diffusion coefficients are fitted within the plotted range for the best match. 
\label{fig:compare2}}
\end{figure}

We then release the constraint of $\mathcal{D}_{ZR} = \mathcal{D}_{RZ} = 0$, allowing them to be non-zero but still constant. Figure \ref{fig:compare2} shows the fitted results in radial and vertical directions. We find the diffusion profiles still do not match the advection ones, even with the contribution from both directions considered. Therefore, a constant diffusion tensor $\overline{\mathcal{D}}$ in the original gradient diffusion hypothesis is not applicable to our models. 

\section{Steps From Equation 22 to Equation 23} \label{app:derivation}

Considering Equation \ref{eq:prhogpt}, \ref{eq:prhodpt}, and \ref{eq:grad-diff-mod}, and multiplying $\langle\rho_{\rm g}\rangle$ to both sides, Equation \ref{eq:lagrangian} writes
\begin{align}
    0 &=  \langle\rho_{\rm g}\rangle\frac{\partial\langle\epsilon_\rho\rangle}{\partial t} 
        + \langle\rho_{\rm g}\rangle\langle\V\rangle\nabla\langle\epsilon_\rho\rangle \\
      &=  \frac{\partial\langle\rho_{\rm d}\rangle}{\partial t} 
        - \langle\epsilon_\rho\rangle\frac{\partial\langle\rho_{\rm g}\rangle}{\partial t}
        + \langle\rho_{\rm g}\rangle\langle\V\rangle\nabla\langle\epsilon_\rho\rangle \\
      &=- \nabla\cdot\bigg(\langle\rho_{\rm d}\rangle\langle\W\rangle\bigg)
        - \nabla\cdot\langle\Delta\rho_{\rm d}\Delta\W\rangle
        + \langle\epsilon_\rho\rangle\nabla\cdot\bigg(\langle\rho_{\rm g}\rangle\langle\V\rangle\bigg)
        + \langle\epsilon_\rho\rangle\nabla\cdot\langle\Delta\rho_{\rm g}\Delta\V\rangle
        + \langle\rho_{\rm g}\rangle\langle\V\rangle\nabla\langle\epsilon_\rho\rangle \\
      &=- \nabla\cdot\bigg(\langle\rho_{\rm d}\rangle\langle\W\rangle\bigg)
        - \nabla\cdot\langle\Delta\rho_{\rm d}\Delta\W\rangle
        + \nabla\cdot\bigg(\langle\epsilon_\rho\rangle\langle\rho_{\rm g}\rangle\langle\V\rangle\bigg)
        + \langle\epsilon_\rho\rangle\nabla\cdot\langle\Delta\rho_{\rm g}\Delta\V\rangle \\
      &=- \nabla\cdot\bigg(\langle\rho_{\rm d}\rangle\langle\W-\V\rangle\bigg)
        - \nabla\cdot\langle\Delta\rho_{\rm d}\Delta\W\rangle
        + \langle\epsilon_\rho\rangle\nabla\cdot\langle\Delta\rho_{\rm g}\Delta\V\rangle \\
      &=- \nabla\cdot\bigg(\langle\rho_{\rm d}\rangle\langle\W-\V\rangle\bigg)
        + \nabla\cdot\bigg(\mathcal{D}\langle\rho_{\rm g}\rangle\nabla\langle\epsilon_\rho\rangle\bigg).
\end{align}